\documentclass[amsmath,amssymb,aps,12pt]{revtex4}
\usepackage{graphicx, bm}
\begin{document}
  \setlength{\baselineskip}{32pt}
  \title{Rhodopsin Photoisomerization:\\Coherent vs. Incoherent Excitation}
  \author{Kunihito Hoki}
  \author{Paul Brumer}
  \affiliation{
    Chemical Physics Theory Group, Department of Chemistry, \\
    and Center for Quantum Information and Quantum  Control,\\
    University of Toronto, Toronto, Canada M5S 3H6 }
\date{May 4, 2006}
  \maketitle
  
  \newpage
  \section*{Abstract}
  A uniform minimal model of rhodopsin photoisomerization induced by
  either coherent laser light or low level incoherent light
  (e.g.~moonlight) is provided. Realistic timescales for both
  processes, which differ by ten orders of magnitude, are obtained.
  Further, a kinetic scheme involving rates for both coherent and
  incoherent light excitation is introduced, placing all timescales
  into a uniform framework. \\
  {\sffamily Keyword}: rhodopsin, isomerization, femtosecond laser,
  incoherent light

  \newpage
  \section{Introduction}
  Developments in fast pulsed lasers have allowed for the detailed
  study of photobiological processes such as laser induced
  {\itshape cis/trans} isomerization of rhodopsin, a process of
  interest due to the large quantum yield ($\sim$ 65\,\%), high speed
  ($\sim$ 200\,fs) of reaction, and importance in the function of
  living organisms~\cite{wang_science_1994, kandori_bc_2001,
    kobayashi_nature_2001}. However, photoinduced processes such as
  this occur naturally in the presence of weak incoherent light,
  rather than in the strong coherent light that emanates from laser
  sources. For example, photoabsorption in rhodopsin initiates
  vertebrate visual transduction in dim light, such as
  moonlight~\cite{sperelakis_1998}. Since the processes induced by
  these two types of sources are qualitatively different, e.g.~pulsed
  coherent light induces time dependent molecular dynamics, whereas
  purely incoherent light does
  not~\cite{jiang_jcp_1991, jiang_cpl_1991}, it is important to
  establish the relationship between them.

  In this paper, we provide a uniform minimal model for
  photoisomerization induced by either of these light sources and
  demonstrate: (a)~a computed dynamics timescale for femtosecond laser
  pulse excitation in agreement with experiment, (b)~realistic
  dynamics for time scales on the order of milliseconds for moonlight
  induced processes, and (c)~a kinetic scheme involving rates of both
  incoherent and coherent excitation that places all timescales within
  a unified framework. Specifically, in the natural visual process,
  the femtosecond coherent timescales provide the initial rise of the
  {\itshape cis/trans} isomerization and the millisecond incoherent
  timescale gives the rate of the process at longer times.

  \section{Theory}
  Our theoretical treatment of the photoisomerization is based on
  a one dimensional system with two electronic states (see
  Fig.~\ref{fig1}a) connecting the {\itshape cis} and {\itshape trans}
  configurations, coupled through a strength parameter $\eta$ to a
  ``bath" that models the effects of the remaining degrees of freedom
  and of the external environment.  Isomerization occurs via rotation about
an  angle $\alpha$. 
The interaction potential between
  the system and the coherent external field $E\!\left(t\right)$ is treated by
  means of the dipole approximation.  In the case of low
  level incoherent light, $E\!\left(t\right)=0$ and a second bath
  describing the incoherent light is included.  That is, our
  Hamiltonian is
  \begin{equation}
    H_{\text{T}}
    = H_{\text{S}} - \mu E\!\left(t\right)
    + H_{\text{Ienv}}  + H_{\text{env}}
    + H_{\text{Irad}} + H_{\text{rad}}, 
  \end{equation}
  where $H_{\text{S}}$ is system Hamiltonian, $\mu$ is transition
  dipole moment of the system, $E\!\left(t\right)$ is electric field
  of the laser pulse, $H_{\text{env}}$ is the environment Hamiltonian,
  $H_{\text{Ienv}}$ is the interaction Hamiltonian between the system
  and environment, $H_{\text{rad}}$ describes blackbody radiation, and
  $H_{\text{Irad}}$ is interaction Hamiltonian between the system and
  the radiation field.  Eigenstates $\left|i\right>$ of the system
  $H_{\text{S}}$ satisfy
  \begin{equation}
    H_{\text{S}} \left|i\right> = \lambda_i \left|i\right>,
  \end{equation}
  and the density matrix accounted with evolution of the (system + bath)
  is denoted $\rho_{\text{T}}$.  The system density matrix is
  $\rho=\text{Tr}_{\text{B}}\rho_T$, where $\text{Tr}_{\text{B}}$
  denotes a trace over the bath.  The time propagation of the density
  matrix elements of the system
  $\rho_{ij}\!\left(t\right)
  =\left<i\right|\rho\!\left(t\right)\left|j\right>$ is
  described by Redfield theory within a secular
  approximation~\cite{redfield_ijrd_1957, blum_plenum_1981,
  may_wiley_2000, pollard_acp_1996} as,
  \begin{align}
    \label{eq:secular_approx1}
    \frac{\partial}{\partial t}\rho_{ii}
    &= \sum_{j \neq i} w_{ij} \rho_{jj}
    - \rho_{ii}\sum_{j \neq i}w_{ji}              \notag \\[12pt]
    &\quad -i\frac{E\!\left(t\right)}{\hbar}\sum_m
    \left[\rho_{im}\!\left(t\right)\mu_{mi}
      -\mu_{im}\rho_{mi}\!\left(t\right)\right]          \\[12pt]
    \label{eq:secular_approx2}
    \frac{\partial}{\partial t}\rho_{ij}
    &= -i\omega_{ij}\rho_{ij}\!\left(t\right)
    -\gamma_{ij}\rho_{ij}\!\left(t\right)         \notag \\[12pt]
    &\quad -i\frac{E\!\left(t\right)}{\hbar}
    \sum_m\left[\rho_{im}\!\left(t\right)\mu_{mj}
      -\mu_{im}\rho_{mj}\!\left(t\right)\right]\qquad (i \neq j),
  \end{align}
  where $w_{ji}=\Gamma^+_{ijji}+\Gamma^-_{ijji}$ is transition
  probability per unit time from $i$th to $j$th eigen state of
  $H_{\text{S}}$, and
  $\gamma_{ij} = \sum_{k}\left(\Gamma^+_{ikki}+\Gamma^-_{jkkj}\right) 
  - \Gamma^+_{jjii} - \Gamma^-_{jjii}$ is dephasing rate.  Here,
  \begin{align}
    \label{eq:gamma}
    \Gamma^+_{ljik}
    &= \frac{1}{\hbar^2} \int_0^{\infty}d\tau e^{-i\omega_{ik}\tau}
    \left<
      H_{\text{Ienv}_{lj}}\!\left(\tau\right) H_{\text{Ienv}_{ik}}
    \right>_{\text{env}} \notag \\[12pt]
    &\quad + \frac{1}{\hbar^2} \int_0^{\infty}d\tau e^{-i\omega_{ik}\tau}
    \left<
      H_{\text{Irad}_{lj}}\!\left(\tau\right) H_{\text{Irad}_{ik}}
    \right>_{\text{rad}} \\[12pt]
    \Gamma^-_{ljik} &= \left(\Gamma^+_{kijl}\right)^*,
  \end{align}
  where the brackets $\left<\dots\right>_{\text{B}}$ represent a trace
  over degrees of freedom in B, where B is either the environment
  ``env'' or the incoherent radiation field ``rad'', and
  $H_{\text{IB}}\!\left(t\right)
  =e^{i H_B t/\hbar}H_{\text{IB}}e^{-i H_B t/\hbar}$.

  The system Hamiltonian $H_{\text{S}}$ is given in terms of two diabatic
  electronic states by
  \begin{equation}
    H_{\text{S}}=\begin{pmatrix}
    T + V_{\text{g}}\!\left(\alpha\right) &
    V_{\text{ge}}\!\left(\alpha\right) \\[12pt]
    V_{\text{eg}}\!\left(\alpha\right) &
    T + V_{\text{e}}\!\left(\alpha\right)
    \end{pmatrix},
  \end{equation}  
  where $T=-\frac{\hbar^2}{2m}\frac{\partial^2}{\partial\alpha^2}$ is
  the kinetic energy, $V_{\text{g}}\!\left(\alpha\right)$ and
  $V_{\text{e}}\!\left(\alpha\right)$ are the potential energy
  surfaces in ground and excited electronic state, and
  $V_{\text{ge}}\!\left(\alpha\right)
  = V_{\text{eg}}\!\left(\alpha\right)$
  is the coupling potential between ground and excited states (see
  Fig.~\ref{fig1}a).

  The environment is described as a set of harmonic oscillators of
  frequency $\omega'_n$ and the system--environment coupling is
  $H_{\text{Ienv}}=Q\sum_n\hbar\kappa_n\left(b_n^{\dagger}+b_n\right)$,
  where $b_n^{\dagger}$ and $b_n$ are the creation and annihilation
  operators pertaining to the $n$th harmonic oscillator.  The operator
  $Q$ is a diagonal $2\times 2$ matrix with $\cos\alpha$ on the
  diagonal, and the coupling constants $\kappa_n$ and spectrum
  of the bath are chosen in accord with an Ohmic spectral density
  $ J\!\left(\omega\right)
  = 2\pi\sum_n\kappa_n^2\delta\!\left(\omega-\omega'_n\right)
  = \eta\omega e^{- \omega / \omega_c}$,
  where the strength of the system--environment coupling is determined
  by the dimensionless parameter $\eta$, and $\omega_c=300$\,cm$^{-1}$.
  After some algebra, we obtain first term of Eq.~\eqref{eq:gamma} as,
  \begin{align}
    \label{eq:gamma1}
    &\frac{1}{\hbar^2} \int_0^{\infty}d\tau e^{-i\omega_{ik}\tau}
    \left<
    H_{\text{Ienv}_{lj}}\!\left(\tau\right) H_{\text{Ienv}_{ik}}
    \right>_{\text{env}} \notag \\[12pt]
    &= \frac{1}{2\pi}Q_{lj}Q_{ik}
    \int_{0}^{\infty}d\tau \int_{0}^{\infty}d\omega
    J\!\left(\omega\right) \cdot \notag \\[12pt]
    & \quad\cdot
    \left\{
    \left[\bar{n}\!\left(\omega\right)+1\right]
    e^{-i\left(\omega_{ik}+\omega\right)\tau}
    + \bar{n}\!\left(\omega\right)
    e^{-i\left(\omega_{ik}-\omega\right)\tau}
    \right\},
  \end{align}
  where $\bar{n}\!\left(\omega\right)
  =\left\{\exp\!\left(\hbar\omega/k_bT\right)-1\right\}^{-1}$ is the
  Bose distribution at temperature $T=300$\,K ,
  $\omega_{ji}=\left(\lambda_j-\lambda_i\right)/\hbar$, and
  $\lambda_i$ is an eigenenergy of $H_{\text{S}}$.

  As a typical situation of scotopic vision, we consider moonlight,
  which is well characterized as a blackbody source at
  4100\,K~\cite{davison_1990}.  The radiation field is also described
  as a set of harmonic oscillators of frequency $\omega''_n$ and the
  system--radiation field coupling is treated by means of dipole
  approximation as,
  \begin{equation}
    H_{\text{Irad}}
    = \mu \sum_{\bm{k}}i\sqrt{\frac{\hbar\omega''_k}{2\epsilon_0 V}}
    \sin\theta
    \left\{
    a_{\bm{k}} \exp\left(i\bm{k}\cdot\bm{r}\right)
    -\smash{a_{\bm{k}}}^{\!\!\dagger}
    \exp\left(-i\bm{k}\cdot\bm{r}\right)
    \right\},
  \end{equation}
  where $\bm{k}$ is a wave number vector, $\epsilon_0$ is the
  permittivity of vacuum, $\bm{r}$ is a position inside of a cavity,
  $V$ is volume of the cavity, and $\theta$ is an angle between the
  transition dipole moment vector and $\bm{k}$~\cite{loudon_1983}.  By
  assuming the large cavity limit the summation of $\bm{k}$ can be
  replaced with integrals, and second term of Eq.~\eqref{eq:gamma} is
  written as,
  \begin{align}
    \label{eq:gamma2}
    &\frac{1}{\hbar^2} \int_0^{\infty}d\tau e^{-i\omega_{ik}\tau}
    \left<
    H_{\text{Irad}_{lj}}\!\left(\tau\right) H_{\text{Irad}_{ik}}
    \right>_{\text{rad}} \notag \\[12pt]
    &= C \frac{\mu_{lj}\mu_{ik}}{2\hbar\epsilon_0\pi^3}
    \int_0^\infty \!d\tau
    \int_0^\infty\!dk
    \int_0^{\frac{\pi}{2}} \!d\theta
    \int_0^{\frac{\pi}{2}} \!d\phi
    k^2 \sin\theta^3 \cdot \notag \\[12pt]
    &\qquad \cdot \left[\left(\bar{n}\!\left(\omega''_k\right) + 1\right)
      e^{-i\left(\omega''_k+\omega_{ik}\right) \tau}
      +\bar{n}\!\left(\omega''_k\right)
      e^{i\left(\omega''_k-\omega_{ik}\right) \tau}
      \right].
  \end{align}
  A component of the imaginary part of Eq.~\eqref{eq:gamma2} describes
  the Lamb shift.  The integration with respect to $k$ does not
  converge, and this difficulty can be avoided by renormalization
  theory~\cite{louisell_wiley_1990}.  However, since the effect of
  Lamb shift is generally less than 0.1\,cm$^{-1}$, the divergent term
  in Eq.~\eqref{eq:gamma2} is neglected in this paper.  The
  coefficient $C$ in Eq.~\eqref{eq:gamma2} is introduced to adjust
  density of blackbody radiation to that of light incident on our
  retina.  Specifically, by assuming that one is looking at a surface lit by
  moonlight, with a color temperature of 4100\,K and a luminance
  $L$\,Cd$\cdot$m$^{-2}$, the ratio of the intensity of light falling
  on the retina over the light falling on the cornea as 0.5, the pupil
  area $3.8 \times 10^{-5}$\,m$^2$, and the distance from the lens to
  the retina of 0.0167\,m, we obtain $C = L/4.0\times 10^{10}$.  Here,
  a conversion from luminous flux in Cd$\cdot$sr to radiant flux in
  W$\cdot$m$^{-1}$ was done by using the spectral luminous efficiency
  function for scotopic vision~\cite{CIE1991}.

  From Eqs.~\eqref{eq:gamma1} and~\eqref{eq:gamma2}, we obtain the
  transition probability in~\eqref{eq:secular_approx1} as,
  \begin{equation}
    w_{ji} = \begin{cases}
      C B_{ji} W\!\left(-\omega_{ji}\right) + A_{ji}
      + \left|Q_{ji}\right|^2 J\!\left(-\omega_{ji}\right)
      \left[\bar{n}\!\left(-\omega_{ji}\right)+1\right]
      &\mspace{13mu}\text{for } \omega_{ji}<0 \\[12pt]
      C B_{ji} W\!\left(\omega_{ji}\right)
      + \left|Q_{ji}\right|^2
      J\!\left(\omega_{ji}\right)\bar{n}\!\left(\omega_{ji}\right)
      &\mspace{13mu}\text{for } \omega_{ji}>0
    \end{cases},
  \end{equation}
  where $A_{ij}$ and $B_{ij}$ are Einstein $A$ and $B$ coefficient in
  between the $i$th and $j$th eigenstate of $H_{\text{S}}$, and
  $W\!\left(\omega\right)$ is the Planck's energy density.  The
  dephasing ratio $\gamma_{ij}$ in~\eqref{eq:secular_approx2} is
  evaluated by numerical integration of Eq.~\eqref{eq:gamma1}.

  \section{Results and Discussion}
  Figure~\ref{fig1}b shows the time propagation of molecular
  populations under a typical laser pulse of time duration 5\,fs,
  amplitude $4\times10^9$\,V/m, and a carrier frequency of
  $2\times10^4$\,cm$^{-1}$ that is resonant with the excitation to the
  electronic excited state around the Franck--Condon region. The
  transition dipole moment, set at 10\,Debye, corresponds to an
  oscillator strength $f\approx 1$. At time $t=0$, the {\itshape cis}
  population $P_{\text{\itshape cis}}\!\left(t=0\right)$ is almost
  unity, and after $t=10$\,fs, probability is created in the excited
  state. Each panel in the Fig.~\ref{fig1}b shows the relaxation
  process with a different degree of system--environment coupling:
  $\eta= 12.5$, 25 and 50. Evident is the fact that the {\itshape
  trans} yield is lower, and the isomerization is faster, with
  increasing coupling $\eta$ to the bath. We note that the time scale
  of the reaction in Fig.~\ref{fig1} is in accord with that observed
  experimentally using coherent light excitation of rhodopsin, i.e. on
  the order of $200$\,fs~\cite{wang_science_1994,kandori_bc_2001}.

  By contrast, the time dependence of the molecular populations for
  the case of excitation by incoherent light is shown in
  Fig.~\ref{fig2}. Here we examine the problem in a context relevant
  to realistic biological systems.  As seen in Fig.~\ref{fig2}, for
  all $\eta$ the rate of 
  increase of $P_{\text{\itshape trans}}$ is linear in time after a
  time that we denote as $t_c (\eta)$. Subsequent to that time the
  slope of $P_{\text{\itshape trans}}$ vs. $t$ is $s = 9.4\times
  10^{-8}$\,s$^{-1}$, corresponding to a {\itshape cis/trans}
  isomerization timescale of almost one year. Note that the slope $s$
  is independent of the speed of photoisomerization observed under
  pulsed laser conditions, as evidenced by the fact that it is 
  independent of $\eta$. Rather, this rate of transformation is
  dictated by the photon flux, which is the rate limiting reagent in
  the process. By contrast, the time $t_c$, which corresponds well to
  the time scale of photoisomerization under the laser pulse, relates
  directly to $\eta$ as $t_c \eta \approx 20\,{\text{ps}}$.  For
  example, for the case of $\eta=12.5$, $t_c= 1.5$\,ps, in accord with
  Fig.~\ref{fig1}b.

  Figure~\ref{fig2}b shows the time dependence of
  $P_{\text{\itshape trans}}$ as a function of the luminance $L$ of
  the incoherent light source. The slope $s$ is seen to be
  proportional to the luminance $L$ as
  $s/L \approx 3.1\times 10^{-6}$\,Cd$^{-1}\cdot$m$^2\cdot$s$^{-1}$.

  Since the isomerization of only a few molecules are necessary to
  induce hyperpolarization in a rod cell~\cite{sperelakis_1998}, we
  compute $P_3$, the probability that at least three from among all of
  the {\itshape cis} molecules in a rod cell are converted to
  {\itshape trans}. The probability would then correspond to the rate
  of our initial visual process under moonlight conditions. The
  probability $P_3(t)$ that at time $t$ at least three from 
  among $N$ molecules are {\itshape trans} is given by
  $1-p_0-p_1-p_2$, where
  \begin{equation}
    p_n=C^n_N p^n (1-p)^{N-n}
  \end{equation}
  is a probability that $n$ from among $N$ molecules are converted to
  {\itshape trans}. Here, $p=P_{\text{\itshape trans}}\!(t)$ is the
  probability that a molecule is {\itshape trans} at time $t$, and
  $C^n_N$ is the binomial coefficient. For the case of vision, we take
  the number of rhodopsin molecules in a rod cell to be
  $N=4 \times$10$^9$~\cite{graymore_1970}, and assume that the time
  dependence of $P_{\text{\itshape trans}}$ maintains a 
  constant slope $s$ until $t = 25\,$msec. The resultant $P_3$ values
  are shown in Fig.~\ref{fig3}, where the time scale to obtain at
  least three {\itshape trans} molecules is on the order of a few tens
  of milliseconds. This finding is consistent with experimental time
  scales of $10\,$msec for dim flash response of a rod
  cell~\cite{sperelakis_1998}. We note, as in the previous results,
  that the speed of photoisomerization under pulsed laser conditions
  bears no relation to the far longer time scales associated with the
  evolution of probability $P_3$, since the photon flux is
  rate-determining in the latter case. Note further that the times at
  which $P_3(t)$ reaches the value of 0.5, a measure of the biological
  response, is virtually a linear function of the irradiance.

  Thus far, molecular time evolution in incoherent light was
  considered using the Redfield approach. We also find that the
  population transfer can be modeled analytically by solving the
  simple three state model with the four reaction rates shown in
  Fig.~\ref{fig4}. A comparison with the computed Fig.~\ref{fig2}
  gives excellent results. Here, states $A$, $B$, and $C$ represent
  {\itshape cis}, excited, and {\itshape trans} conformations of the
  molecule, respectively. The values of $k_2$ and $k_4$ correspond to
  rates of population transfer from $P_{\text{e}}$ to
  $P_{\text{\itshape cis}}$ and $P_{\text{\itshape trans}}$, which are
  mainly caused by the system--environment coupling. Values obtained
  from the coherent pulse studies of Fig.~\ref{fig1} give
  $k_2 = k_4 = 0.08 \eta$\,ps$^{-1}$. The $k_1$ and $k_3$ represent
  rates of population transfer from $P_{\text{\itshape cis}}$ and
  $P_{\text{\itshape trans}}$ to $P_{\text{e}}$, caused by both
  system--environment coupling and photoabsorption. The rates of
  system--environment coupling can be assigned using detailed balance,
  and the rates of photoabsorption are given by the Einstein
  transition probability from the electronic ground state to the
  electronic excited state. In the case of $k_1$, the primary
  contribution is photoabsorption, giving $k_1 = BW = L \times 5.6
  \times10^{-6}$\,Cd$^{-1}\cdot$m$^2\cdot$s$^{-1}$, where $B$ is the
  Einstein $B$ coefficient, and $W$ is density of energy of the
  radiation field. The densities of the field used in Fig.~\ref{fig4}
  correspond to the luminescence values used in
  Fig.~\ref{fig2}~\cite{note2}. On the other hand, in the case of
  $k_3$, the dominant term is system--environment coupling, and we
  obtain $k_3 = k_4\times 1.87\times 10^{-9}$. With the resultant
  $k_1$,~$k_3 << k_2$,~$k_4$, the rate equations give the reaction
  rate for isomerization under incoherent light as $k_1/2 =
  BW/2$. Further, these equations establish the existence of a linear
  region for $P_{\text{\itshape trans}}$ vs.~$t$ with an $\eta$
  independent slope $s=k_1/2$ after a time $t_c=3/(k_2 + k_4)$,
  relating the rate approach to both the computed coherent and
  incoherent results. 

  We note that the reaction rate obtained by the
  three state model is $\approx$10\,\% smaller than that given by the
  Redfield equation. The difference mainly comes from the simplifying
  assumption that $k_2 = k_4$, and the evaluation of the rate of 
  photoabsorption  at the torsional angle $\alpha$ set to zero.  
  Nonetheless, all of the trends seen in
  the Redfield computed results are also evident in the rate equation
  results.

  \section{Summary}
  We have presented a unified theoretical model of photoisomerization
  under both a coherent light source such as a femtosecond laser pulse
  and an incoherent light source such as the moonlight. A minimal
  model of the isomerization process that gives the same timescale as
  the femtosecond laser experiment was obtained. It was shown that the
  time scale for photoisomerization under coherent light corresponds
  to the initial rise time $t_c$ of the photoisomerization under
  incoherent light. Further, we introduced a simple three state model
  that incorporates all of the relevant rates obtained from both the
  femtosecond and millisecond time domains.

  This approach provides a connection between the time domain of the
  femtosecond laser experiment and that of biologically relevant
  response time scales. A dynamical behavior is seen even for
  the case of incoherent light source, since sudden irradiation of
  the light at $t=0$ introduces partial coherence into the system.
  The very earliest dynamics correlate with the primary event of
  isomerization as identified in femtosecond laser experiments. The
  exact response observed reflects the combined effect of the
  characteristics of the radiation field and the underlying
  dynamics. 

  Acknowledgment: This work was carried out with partial support from
  Photonics Research Ontario and NSERC Canada. We thank Professor
  R.J.~Dwayne Miller for extensive comments on a earlier version of
  this manuscript.

  \newpage
  \bibliography{paper}

\begin{thebibliography}{16}
\bibitem[{not()}]{note2}
\bibinfo{note}{Here $W = 2.5\times 10^{-11} w(\omega) L$, where the coefficient
  is estimated by using the radius of pupil, the distance between the lens and
  a surface, etc., $L$ is luminance of the surface in Cd$\cdot$m$^{-2}$, and
  $w(\omega)$ is the black body energy density at 4100K. Here, $W$ and
  $w(\omega)$ have units of Joule$\cdot$sec$\cdot$m$^{-3}$.}
\end{thebibliography}

  \newpage
  \section*{Figure Captions}
\begin{figure}[htbp]
    \begin{center}
      \includegraphics[width=6cm]{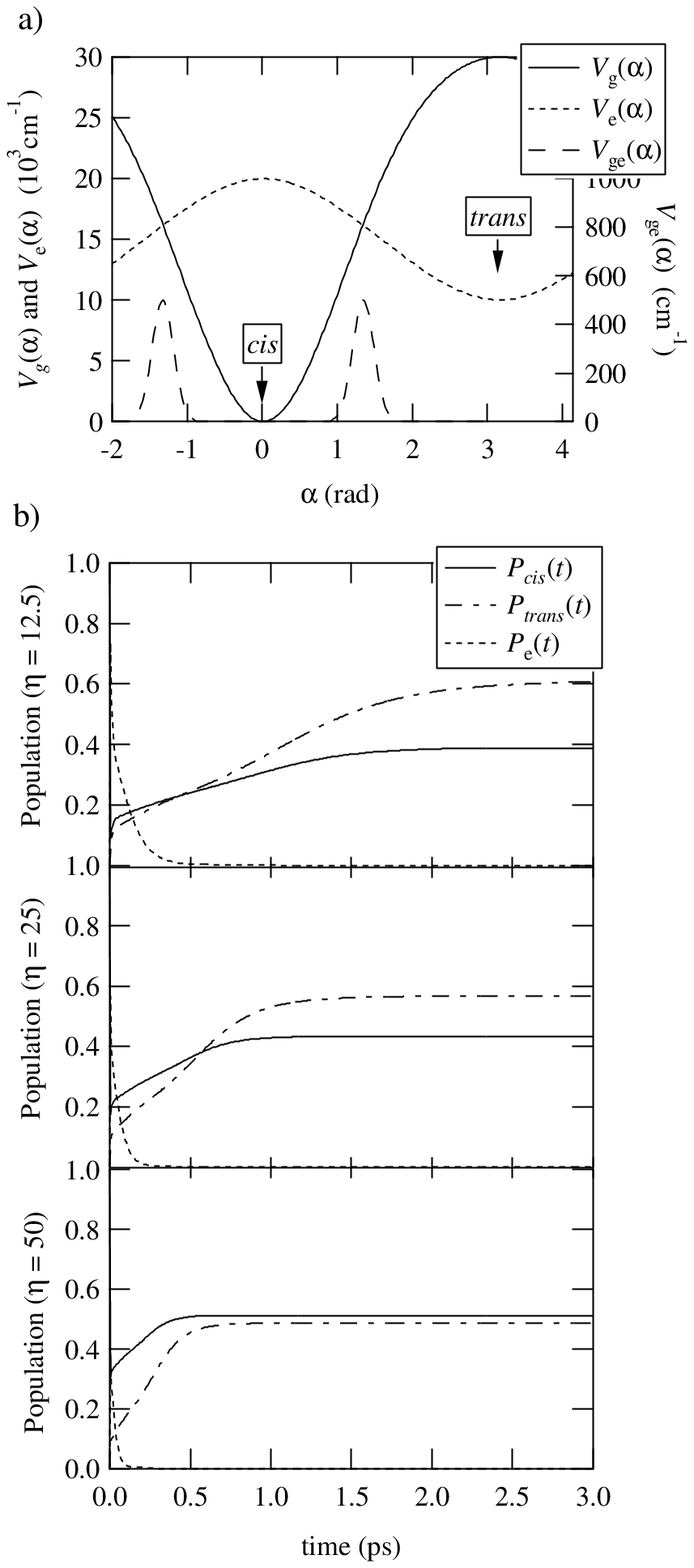}
    \end{center}
    \caption{}
    \label{fig1}
  \end{figure}

  FIG.~\ref{fig1}: a) Potential energy surfaces for the two state model for
  {\itshape cis} to {\itshape trans} photoisomerization. The solid
  curve and dotted curve show diabatic potentials $V_{\text{g}}$ and
  $V_{e}$, respectively. The dashed curve shows a coupling potential
  between two diabatic electronic states. b)~Time propagation of cis
  and trans populations under a short intense pulse for different
  values of $\eta$. $P_{\text{\itshape cis}}$ is the population in the
  range $-\frac{\pi}{3}\leq\alpha\leq\frac{\pi}{3}$ on $V_{\text{g}}$,
  $P_{\text{\itshape trans}}$ is that in the range
  $-\pi\leq\alpha\leq-\frac{2\pi}{3}$ on $V_{\text{e}}$, and
  $P_{\text{e}}=1-P_{\text{\itshape cis}}-P_{\text{\itshape trans}}$.
  Note that in Panels b and c, the very short time dynamics, which
  includes the excitation from the {\itshape cis}, is not evident
  due to the short time over which it occurs.

\begin{figure}[htbp]
    \begin{center}
      \includegraphics[width=9cm]{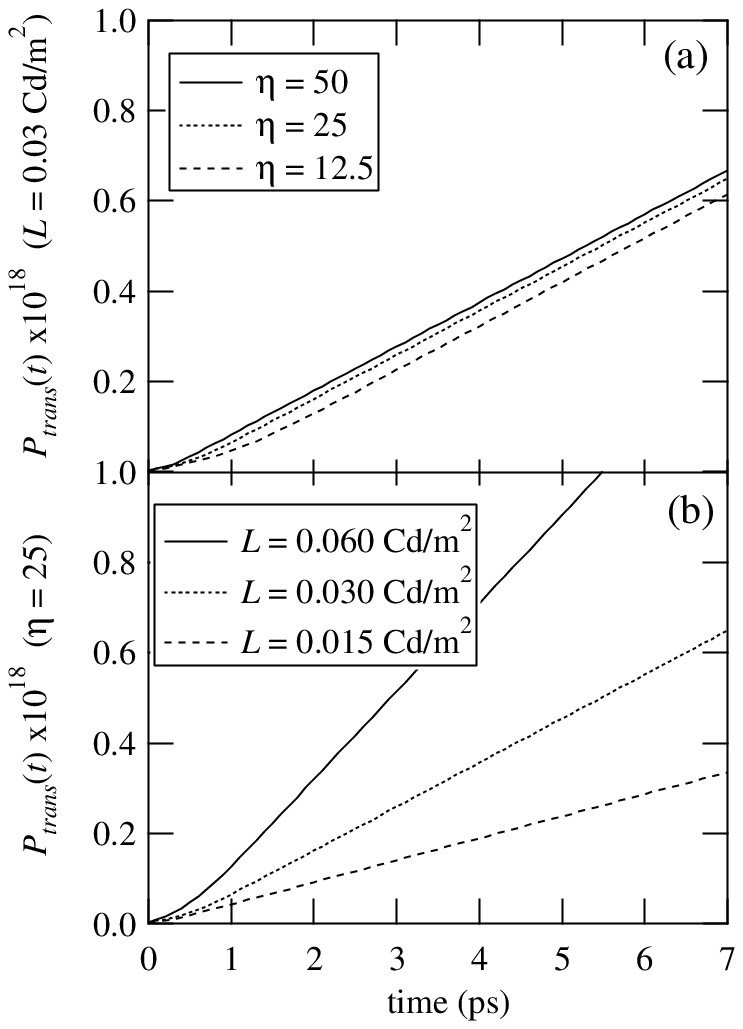}
    \end{center}
    \caption{}
    \label{fig2}
  \end{figure}

  FIG.~\ref{fig2}: a) Time dependence of $P_{\text{\itshape trans}}$
  for three $\eta$ values with incoherent light luminescence
  $L=0.03$\,Cd$\cdot$m$^{-2}$. b)~Time dependence of
  $P_{\text{\itshape trans}}$ for various values of $L$, with
  system--environment coupling $\eta = 25$. In all cases there is a
  deviation from strictly linear behavior at the early times that
  corresponds to timescales of isomerization dynamics.

\begin{figure}[htbp]
    \begin{center}
      \includegraphics[width=9cm]{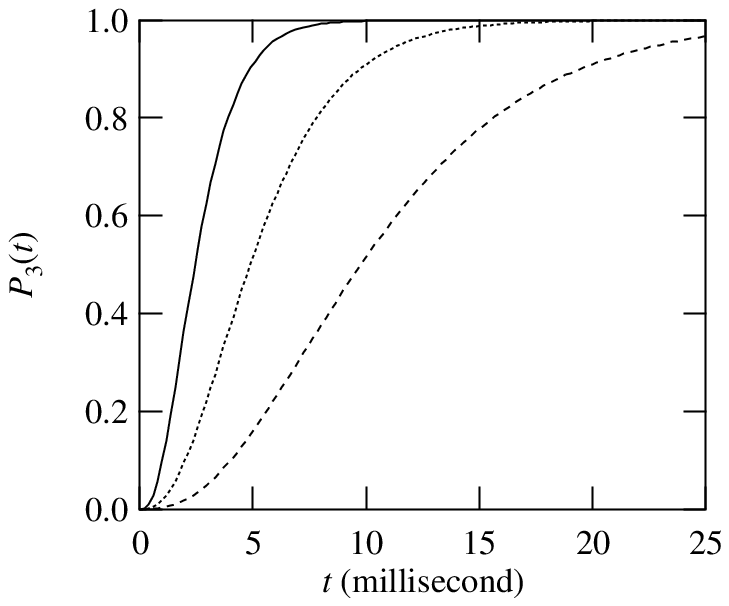}
    \end{center}
    \caption{} 
    \label{fig3}
  \end{figure}

  FIG.~\ref{fig3}: Time dependence of the probability $P_3$ that at
  least three from among $4 \times$10$^9$ {\itshape cis} molecules
  become {\itshape trans} for various values of $L$: 0.060\,Cd/m$^2$
  (solid); 0.030\,Cd/m$^2$  (dotted); 0.015\,Cd/m$^2$ (dashed).

\begin{figure}[htbp]
    \begin{center}
      \includegraphics[width=7cm]{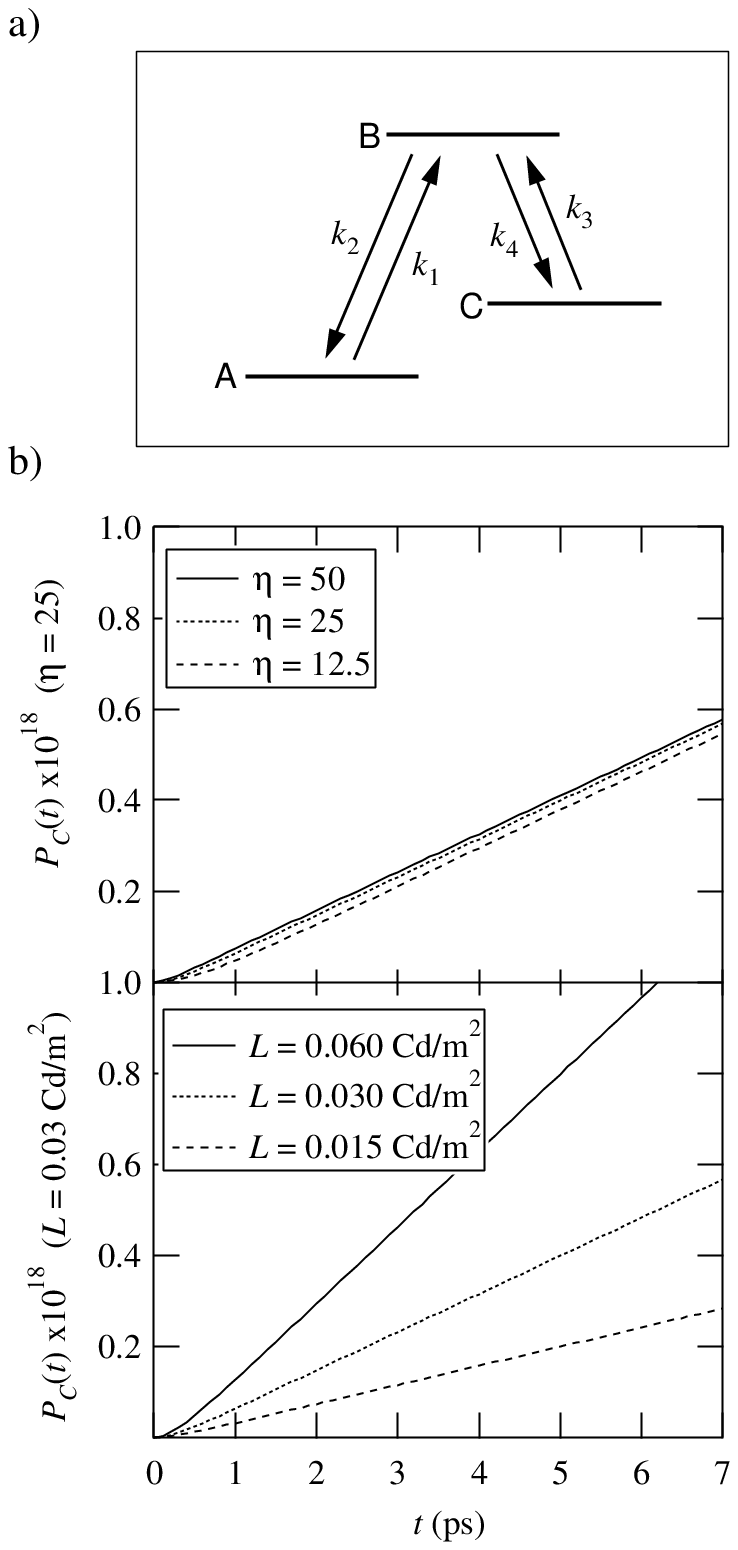}
    \end{center}
    \caption{}
    \label{fig4}
  \end{figure}

  FIG.~\ref{fig4}: a) Three states model with reaction rates $k_1$,
  $k_2$, $k_3$, and $k_4$. Time propagation of $P_C$ for each
  luminance $L$ and strength parameter of system--environment coupling
  $\eta$.  Compare with results shown in Fig.~\ref{fig2}.

  \newpage

  \newpage
  
  \newpage

  \newpage
\end{document}